\documentclass[twocolumn,showpacs,preprintnumbers,superscriptaddress,prb]{revtex4}
\usepackage{graphicx,bm,times,url}
\usepackage{amsmath}
\graphicspath{{./fig/}{./png/}}
\newcommand{\BoldVec}[1]{\mathchoice%
  {\mbox{\boldmath $\displaystyle     #1$}}%
  {\mbox{\boldmath $\textstyle        #1$}}%
  {\mbox{\boldmath $\scriptstyle      #1$}}%
  {\mbox{\boldmath $\scriptscriptstyle#1$}}%
}
\newcommand{\EQ}{\begin{equation}}
\newcommand{\EN}{\end{equation}}
\newcommand{\EQA}{\begin{eqnarray}}
\newcommand{\ENA}{\end{eqnarray}}
\newcommand{\eq}[1]{(\ref{#1})}
\newcommand{\EEq}[1]{Equation~(\ref{#1})}
\newcommand{\Eq}[1]{Eq.~(\ref{#1})}
\newcommand{\Eqs}[2]{Eqs.~(\ref{#1}) and~(\ref{#2})}

\newcommand{\eqs}[2]{(\ref{#1}) and~(\ref{#2})}

\newcommand{\Sec}[1]{\S\,\ref{#1}}

\newcommand{\App}[1]{Appendix~\ref{#1}}
\newcommand{\Fig}[1]{Fig.~\ref{#1}}

\newcommand{\Tab}[1]{Table~\ref{#1}}
\newcommand{\Figs}[2]{Figs.~\ref{#1} and \ref{#2}}

\newcommand{\meanBB}{\overline{\bm{B}}}

\newcommand{\meanEMF}{\overline{\mbox{\boldmath ${\cal E}$}} {}}

\newcommand{\hr}{\mbox{$h^{\rm r}$}}
\newcommand{\ha}{\mbox{$h^{\rm a}$}}
\newcommand{\har}{\mbox{$h^{\rm ar}$}}

\newcommand{\hf}{\overline{h}_{\rm f}}
\newcommand{\kf}{\mbox{$k_{\rm f}$}}
\newcommand{\uu}{\BoldVec{u} {}}

\newcommand{\UU}{\BoldVec{U} {}}

\newcommand{\bb}{\BoldVec{b} {}}

\newcommand{\BB}{\BoldVec{B} {}}

\newcommand{\AAA}{\BoldVec{A} {}}

\newcommand{\AAAW}{\BoldVec{A}^{\rm W}}
\newcommand{\AAAr}{\BoldVec{A}^{\rm r}}
\newcommand{\AAAa}{\BoldVec{A}^{\rm a}}
\newcommand{\AAAar}{\BoldVec{A}^{\rm ar}}
\newcommand{\FFFr}{\BoldVec{F}^{\rm r}}
\newcommand{\FFFa}{\BoldVec{F}^{\rm a}}
\newcommand{\FFFar}{\BoldVec{F}^{\rm ar}}
\newcommand{\FFFf}{\overline{\BoldVec{F}}_{\rm f}}
\newcommand{\aaa}{\BoldVec{a} {}}

\newcommand{\jj}{\BoldVec{j} {}}
\newcommand{\JJ}{\BoldVec{J} {}}

\newcommand{\ff}{\BoldVec{f} {}}

\newcommand{\FF}{\BoldVec{F} {}}

\newcommand{\nab}{\BoldVec{\nabla} {}}

\newcommand{\AW}{A^{\rm W}}
\newcommand{\Aa}{A^{\rm a}}
\newcommand{\Ar}{A^{\rm r}}
\newcommand{\Aar}{A^{\rm ar}}

\newcommand{\LamWa}{\Lambda^{\rm W:a}}
\newcommand{\Lamrar}{\Lambda^{\rm r:ar}}

\newcommand{\SSSS}{\bm{\mathsf{S}}}

\newcommand{\DD}{{\rm D} {}}
\newcommand{\dd}{{\rm d} {}}

\newcommand{\const}{{\rm const}  {}}

\def\Sc{\mbox{\rm Sc}}
\def\Pm{\mbox{\rm Pr}_{\rm M}}
\def\Rm{\mbox{\rm Re}_{\rm M}}
\def\Rey{\mbox{\rm Re}}

\def\cs{c_{\rm s}}
\def\kf{k_{\rm f}}
\def\urms{u_{\rm rms}}

\begin{document}
\preprint{NORDITA 2010-94}

\title{Magnetic helicity transport in the advective gauge family}

\author{Simon Candelaresi}
\affiliation{NORDITA, AlbaNova University Center, Roslagstullsbacken 23,
SE-10691 Stockholm, Sweden}
\affiliation{Department of Astronomy,
Stockholm University, SE 10691 Stockholm, Sweden}

\author{Alexander Hubbard}
\affiliation{NORDITA, AlbaNova University Center, Roslagstullsbacken 23,
SE-10691 Stockholm, Sweden}

\author{Axel Brandenburg}
\affiliation{NORDITA, AlbaNova University Center, Roslagstullsbacken 23,
SE-10691 Stockholm, Sweden}
\affiliation{Department of Astronomy,
Stockholm University, SE 10691 Stockholm, Sweden}

\author{Dhrubaditya Mitra}
\affiliation{NORDITA, AlbaNova University Center, Roslagstullsbacken 23,
SE-10691 Stockholm, Sweden}

\date{\today}

\begin{abstract}
Magnetic helicity fluxes are investigated in a family of gauges in which
the contribution from ideal magnetohydrodynamics takes the form of a
purely advective flux.
Numerical simulations of magnetohydrodynamic turbulence
in this advective gauge family exhibit instabilities triggered by
the build-up of unphysical irrotational contributions to the
magnetic vector potential.
As a remedy, the vector potential is evolved
in a numerically well behaved gauge, from which 
the advective vector potential is obtained by a gauge transformation.
In the kinematic regime, the magnetic helicity density evolves similarly
to a passive scalar when resistivity is small and turbulent mixing is mild,
i.e.\ when the fluid Reynolds number is not too large.
In the dynamical regime, resistive contributions to the magnetic helicity
flux in the advective gauge
are found to be significant owing to the development of small length
scales in the irrotational part of the magnetic vector potential.
\end{abstract}


\maketitle

\section{Introduction}
\label{Introduction}
Most astrophysical and laboratory plasmas are good conductors.
This, together with high-speed flows and large length scales,
nearly universal in the astrophysical context, makes for large magnetic
Reynolds numbers.
In the limit of infinitely large magnetic Reynolds number, and 
for domains with closed boundaries, total magnetic helicity is a
conserved quantity. 
Here, an analogy can be drawn with mass conservation in domains whose
boundaries are closed to mass flux. 
Furthermore, in open domains, the change in total mass
is governed by the mass flux across open surfaces.
In ideal magnetohydrodynamics (MHD), a similar property holds
for the total magnetic helicity.
But unlike mass, magnetic helicity depends on the choice of gauge.
In the special case of the advective gauge, the magnetic helicity
flux is given by the velocity times the magnetic helicity density
\cite{HubbardBrandenburg2010ApJ}, making this gauge particularly
interesting for studying pointwise properties of magnetic helicity.
This is an important goal of this paper.

Magnetic helicity plays an important role in many fields of
plasma physics and astrophysics, and has applications ranging
from tokamaks and other plasma confinement machines, to dynamo
action in the Sun and the Galaxy.
Our physical understanding of the role of magnetic helicity in MHD
is greatly aided by concepts such as Taylor relaxation \cite{Taylor1986},
selective decay \cite{MM80},
and the inverse cascade of magnetic helicity \cite{PFL}.

Furthermore, magnetic helicity is a crucial ingredient of 
the turbulent dynamos which are believed to be the source of the
equipartition magnetic fields in astrophysical bodies like stars and 
galaxies \citep{BrandenburgSubramanian2005PhR}.
In all such cases the characteristic 
length scales of the dynamo generated magnetic field exceed those of the
fluid's energy carrying scale.  
In dynamo theory, the formation of such a large-scale magnetic field
is typically possible through the $\alpha$ effect,
which is non-zero for helical turbulent flows.
In periodic boxes with helical turbulence, the $\alpha$ effect
becomes strongly quenched when the (appropriately normalized) magnetic helicity
in the small-scale field
(i.e., scales that are smaller than the energy-carrying scale of turbulent
fluid) 
is comparable to the helicity in the small-scale velocity.
Conservation of magnetic helicity implies that the helicity in small-
and large-scale fields will have comparable magnitudes, so the quenching
of the large-scale dynamo will occur for weak large-scale fields.
This $\alpha$ quenching \cite{VainshteinCattaneo1992, CattaneoHughes1996}
increases with scale separation and endures for as long as magnetic
helicity is nearly conserved, a resistive time that scales with the
magnetic Reynolds number $\Rm \equiv UL/\eta$.
The quenching is called
``cata\-strophic'' because for the Sun $\Rm \sim 10^{9}$ and the Galaxy
$\Rm \sim 10^{15}$, and their resistive timescales are problematically long.
This rapid pre-resistive saturation of the dynamo generated field poses clear
difficulties in applying theory to astronomical systems,
but it may be possible to alleviate the problem through magnetic helicity fluxes
\cite{BlackmanField2000MNRAS,BlackmanField2000ApJ}.
It should also be pointed out that problems with catastrophic quenching
are often not clearly seen in present-day simulations
\cite{Brown10,Kapy10,Charb10}.  While trend lines suggest that
catastrophic quenching will occur, simulations at currently achievable, 
low to intermediate $\Rm$ and scale separation
have shown significant large-scale fields.

There exists reasonable observational evidence is support of such
fluxes of magnetic helicity.
The Sun's surface magnetic field shows helical structures \cite{Gibson2002,Canfield1996b}.
Further, it was shown \cite{Manoharan1996} that the S-shaped (helical)
regions which are active in the corona are precursors of coronal
mass ejections (CMEs) and later \cite{Canfield1999} that those regions are more
likely to erupt.
This suggests that the Sun sheds magnetic helicity via CMEs.
Since the Sun's large-scale magnetic field is believed to be
generated by a helical dynamo \cite{Shukurov2006,ssHelLoss09}
this shedding of magnetic helicity could play an
important role in the 11 year solar cycle.
Physically, magnetic helicity fluxes out of the domain can be mediated
in many ways, such as the aforementioned CMEs
for the Sun \cite{BlackmanBrandenburgApJ2003} or
fountain flows in the case of galaxies \cite{Shukurov2006}.
In direct simulations magnetic helicity fluxes are permitted
by adjusting the boundary conditions, e.g., to vertical field
boundaries, but their actual presence can be difficult to ascertain. 
Internal helicity fluxes have also been found to
alleviate $\alpha$ quenching \cite{ssHelLoss09} in systems
with internal boundaries 
that separate zones of oppositely signed kinetic and magnetic helicities.

A difficulty in addressing the generation and transport of magnetic helicity
is its gauge dependence.  We denote the magnetic vector potential as $\AAA$
such that $\BB \equiv \nab\times\AAA$ is the magnetic field.  Magnetic helicity
$H \equiv \int_{V} \AAA\cdot\BB \ \dd V$ is independent of the gauge 
for perfectly conducting boundaries, as well as periodic boundaries
so long as $\AAA$ is also required to be periodic.
However, if one wishes to study the
transport of magnetic helicity for physically motived systems
a non-volume integral formulation will be needed. 
Magnetic helicity \emph{density}, $h \equiv \AAA\cdot\BB$,
 the quantity we will be working with, clearly depends on the gauge
choice for $\AAA$.  The gauge dependence of fluxes of
mean magnetic helicity contained in the fluctuating fields
was examined via direct numerical simulations (DNS)
for three different gauges \cite{helGauge10}, and it
was found that, averaged over time, they do not depend on the gauge choice.
This is a result of the fact that, for sufficient scale separation, the
magnetic helicity of the fluctuating field can be expressed as the density
of linkages, which in turn is gauge-invariant \cite{SB06}.
This result implies that the study of specific but useful gauge choices is
a meaningful task. 

In this work we examine the properties of magnetic helicity density in a
particularly interesting
gauge-family which we call ``advective" because in this gauge the effect of 
velocity on the evolution equation of magnetic helicity takes the form of
a purely advective term.
In previous work\cite{HubbardBrandenburg2010ApJ} this gauge choice was shown to
be crucial to understanding magnetic helicity fluxes in the presence of shear,
including the Vishniac--Cho flux \cite{VishniacCho2001ApJ}.
Unfortunately, evolving $\AAA$ in this gauge proves numerically unstable.
This may be related to earlier findings in smoothed particle
MHD calculations \cite{Price09,Price10}.
There, the problem was identified as the result of an
unconstrained evolution of vector potential components, which was argued to be
connected with ``poor accuracy with
respect to `reverse-advection'-type terms'' \cite{Price09}.
Our present work clarifies that this instability is related to
the excessive build-up of irrotational contributions to the magnetic
vector potential.
These contributions have no physical meaning, but discretization errors
at small length scales can spoil the solution dramatically.

We shall therefore describe a novel method for obtaining $\AAA$ in this
gauge by evolving it first in a numerically robust gauge and then applying
a gauge transformation with a simultaneously evolved gauge potential.
This will be referred to as the $\Lambda$ method throughout the text.
Next, we show that the magnetic helicity density in the advective gauge
tends to be small even pointwise, provided turbulent effects are still
weak, and discuss the analogy with passive scalar transport.
We conclude by pointing out that resistive terms break the analogy
with passive scalar advection through the emergence of a turbulently
diffusive magnetic helicity flux.

\section{Magnetic evolution equations}

\subsection{Weyl and advective gauges}
\label{Model}

In this work we remain within non-relativistic MHD and hence neglect the 
Faraday displacement current. So the current
density is given by $\JJ=\nab\times\BB$, where $\BB$ is the magnetic field
and we use units where the vacuum permeability is unity.
At the core of MHD is the induction equation,
\EQ
\frac{\partial \BB}{\partial t}
=\nab\times(\UU \times \BB - \eta \JJ),
\label{induction1}
\EN
where $\UU$ is the velocity and $\eta$ is the molecular magnetic diffusivity.
\EEq{induction1} can be uncurled to give an evolution equation for
the magnetic vector potential $\AAA$, but only up to a gauge choice.
In the Weyl gauge, indicated by a superscript W on the magnetic
vector potential, we just have
\EQ
\frac{\partial\AAAW}{\partial t}=\UU \times \BB - \eta \JJ,
\label{Weyl}
\EN
but by adding the gradient of a scalar field, the vector potential can
be obtained in any other gauge.
Of particular interest to this paper is the advective gauge,
\EQ
\AAAa=\AAAW+\nab\LamWa,
\label{LamWa}
\EN
where $\LamWa$ is the gauge potential that transforms from $\AAAW$ to $\AAAa$.
We demand that\cite{B10}
\EQ
{{\DD\Aa_i}\over\DD t}=-U_{j,i}\Aa_j-\eta J_{i}.
\label{dAAAa}
\EN
Here, $\DD/\DD t=\partial/\partial t+\UU\cdot\nab$ is the advective
derivative.
Consequently one can show that $\LamWa$ obeys the evolution equation
(see \App{sec: Derivation of Lambda})
\EQ
{\DD\LamWa\over\DD t}=-\UU\cdot\AAAW.
\label{dLamWa}
\EN
Thus, to obtain $\AAAa$, one can either solve \Eq{dAAAa} directly
or, alternatively, solve \Eq{Weyl} together with \Eq{dLamWa} and
use \Eq{LamWa} to obtain $\AAAa$.
A possible initial condition for $\LamWa$ would be $\LamWa=0$,
in which case $\AAAa=\AAAW$ initially.
For numerical reasons that will be discussed in more detail below,
we shall consider the indirect method of obtaining the magnetic vector
potential in the advective gauge, but starting from more numerically stable 
gauge which will be discussed in \Sec{ResistiveAdvRes}.

Variants on the advective gauge have seen significant use, particularly
in DNS with constant imposed shear.
Although the magnetic field in such simulations must obey
shearing-periodic boundary condition the vector potential need not. 
In particular, the evolution equation (\ref{Weyl}) does
not impose shearing-periodicity on the vector potential, while \Eq{dAAAa} does,
enabling shearing-periodic numerical simulations \citep{BNST95} in terms
of $\AAA$.

For our purposes, the importance of \Eq{dAAAa} lies in the form of the
magnetic helicity density evolution equation.
By writing the induction equation in the form
\EQ
{{\DD B_i}\over\DD t}=+U_{i,j}B_j-(\nab\cdot\UU)B_i-(\nab\times\eta\JJ)_i,
\EN
computing
$\DD(\AAAa\cdot\BB)/\DD t=\AAAa\cdot\DD\BB/\DD t+\BB\cdot\DD\AAAa/\DD t$,
and noting that the $A_i U_{i,j} B_j$ terms from both equations cancel,
we find that
\EQ
{\DD\ha\over\DD t}=-\ha\nab\cdot\UU
-\nab\cdot(\eta\JJ\times\AAAa)
-2\eta\JJ\cdot\BB,
\label{h1}
\EN
which shows that in ideal MHD ($\eta=0$)
under the assumption of incompressibility ($\nab\cdot\UU=0$)
the magnetic helicity density in the advective gauge, $\ha=\AAAa\cdot\BB$
is just advected with the flow like a passive scalar, i.e.\
\EQ
{\DD\ha\over\DD t}=0\quad\mbox{(for $\eta=0$ and $\nab\cdot\UU=0$)}.
\label{h1b}
\EN
In the general case with $\nab\cdot\UU\neq0$, the rate of change of
the local value of $\ha$ is given by $-\nab\cdot(\ha\UU)$, which
is analogous to the continuity equation for the fluid density.
However, for $\eta\neq0$, there is also a source term,
\EQ
{\partial\ha\over\partial t}=-2\eta\JJ\cdot\BB-\nab\cdot\FFFa,
\label{h1c}
\EN
as well as a resistive contribution to the magnetic helicity flux,
\EQ
\FFFa=\ha\UU+\eta\JJ\times\AAAa.
\label{h1flux}
\EN
In this paper we address the question how the $\eta\JJ\times\AAAa$
contribution scales in the limit $\eta\to0$, i.e.\ for large values
of $\Rm$.
It could either stay finite, just like the resistive energy
dissipation $\eta \JJ^2$, which
tends to a finite limit \cite{BrandenburgSubramanian2005PhR} as $\eta\to0$,
or it could go to zero like the source term
$\eta \JJ \cdot \BB$ \cite{Ber84,Brandenburg2001ApJ}.

\subsection{Resistive and advecto-resistive gauges}
\label{ResistiveAdvRes}
There are two important issues to be noted about the equations discussed above.
Firstly, for numerical reasons, \Eq{Weyl} is often replaced by
\EQ
\frac{\partial \AAAr}{\partial t}=\UU \times \BB +\eta \nabla^2\AAAr,
\label{rdef1}
\EN
where $\AAAr$ is the magnetic vector potential in the resistive gauge and
we have assumed that $\eta=\const$; otherwise there would be an
additional gradient term of the magnetic diffusivity
that results from \cite{DSB02}
\EQ
-\eta\JJ+\nab(\eta\nab\cdot\AAA)=\eta\nabla^2\AAA+(\nab\cdot\AAA)\nab\eta.
\EN
This ``resistive'' gauge introduces an explicit, numerically stabilizing
diffusion term for each component of $\AAA$.
Secondly, and again for numerical reasons, \Eq{dLamWa} should be solved
with a small diffusion term proportional to $\nabla^2\LamWa$.
These two issues are actually connected and can be resolved by considering
the gauge transformation
\EQ
\AAAar=\AAAr+\nab\Lamrar,
\label{Lamrar}
\EN
which allows us to obtain the magnetic vector potential $\AAAar$ in the
advecto-resistive gauge obeying
\EQ
{{\DD\Aar_i}\over\DD t}=-U_{j,i}\Aar_j+\eta\nabla^2\Aar_i,
\label{dAAAar}
\EN
by solving \Eq{rdef1} for $\AAAr$ together with
\EQ
{\DD\Lamrar\over\DD t}=-\UU\cdot\AAAr+\eta\nabla^2\Lamrar
\label{dLdt1}
\EN
and finally using the gauge transformation \Eq{Lamrar}.
For a full derivation of this equation we refer to
\App{sec: Derivation of dLdt1}.
Note that the microscopic magnetic diffusivity automatically enters
the $\Lamrar$ equation as a diffusion term,
which implies that the $\Lamrar$ equation is numerically well behaved.

The magnetic helicity density $\har=\AAAar\cdot\BB$ in the
advecto-resistive gauge can be calculated from the magnetic
helicity in the resistive gauge through $\har=\hr+\nab\Lamrar\cdot\BB$,
and it obeys
\EQ
{\partial\har\over\partial t}=-2\eta\JJ\cdot\BB-\nab\cdot\FFFar
\label{dhar}
\EN
with
\EQ
\FFFar=\har\UU-\eta(\nab\cdot\AAAar)\BB+\eta\JJ\times\AAAar.
\label{FFFar}
\EN
For comparison, the evolution equation of the magnetic helicity density
in the resistive gauge is given by an equation similar to \eq{dhar},
but with $\har$ being replaced by $\hr$ and $\FFFar$ being replaced by
\EQ
\FFFr=\hr\UU-(\UU\cdot\AAAr+\eta\nab\cdot\AAAr)\BB+\eta\JJ\times\AAAr,
\label{non-adv}
\EN
which contains a non-advective velocity driven flux
of the form $(\UU\cdot\AAAr)\BB$ -- even in the ideal case.

\subsection{Numerical details}

We perform simulations for isotropically forced, triply periodic cubic domains 
with sides of length $2\pi$, as was done in earlier work
\cite{Brandenburg2001ApJ}.
The $\eta \JJ \cdot \BB$ term in \eq{h1c} implies
(and past simulations have shown) that such a system will experience a
slow, but steady
production of magnetic helicity.  This is the price to pay for a system
which is both helical, providing us with a signal, and homogeneous, so avoiding
extraneous magnetic helicity fluxes.
In addition to the uncurled induction equation \eq{rdef1} and
the gauge transformation evolution equation \eq{dLdt1}, we solve
\begin{eqnarray}
&&\frac{\DD\UU}{\DD t} =  -\cs^{2}\nab\ln{\rho} + \frac{c_{\rm L}}{\rho}\JJ\times\BB
 + \FF_{\rm visc} + \ff, \\
&&\frac{\DD\ln\rho}{\DD t}=  -\nab\cdot\UU,
\end{eqnarray}
where $\cs$ ($=\const$) is the isothermal sound speed, $\rho$ is the density,
$\FF_{\rm visc} = \rho^{-1}\nab\cdot(2\rho\nu\SSSS)$ is the viscous force,
${\sf S}_{ij}=\frac{1}{2}(U_{i,j}+U_{j,i})-\frac{1}{3}\delta_{ij}\nab\cdot\UU$
is the rate of strain tensor, $\nu$ is the kinematic viscosity,
$\ff$ the forcing term, and $c_{\rm L}=1$ is a prefactor that can
be put to $0$ to turn off the Lorentz force in kinematic calculations.
As in earlier work \cite{Brandenburg2001ApJ} the forcing function consists
of plane polarized waves whose direction and phase change randomly from
one time step to the next.
The modulus of its wavevectors is taken from a band of wavenumbers
around a given average wavenumber $k_{\rm f}$.
The magnetic vector potential is initialized with a weak non-helical
sine wave along one direction.
In some cases we shall also consider solutions to the passive scalar
equation in the incompressible case,
\EQ
{\DD C\over\DD t}=\kappa\nabla^2C,
\EN
where $\kappa$ is the passive scalar diffusivity.
Following earlier work \cite{BKM04}, we impose a linear gradient in $C$,
i.e.\ $C=Gz+c$, and solve for the departure from this gradient $G$, i.e.\
\EQ
{\DD c\over\DD t}=\kappa\nabla^2c-GU_z,
\EN
where $GU_z$ acts essentially as a forcing term.

We use the {\sc Pencil Code} (http://pencil-code.googlecode.com)
\cite{PencilCode} to solve
the equations for $\AAAr$, $\UU$, $\Lamrar$, $\rho$,
and in some cases also $c$.
The calculations involving $\Lamrar$ have been carried out with the
publicly available revision \url{r15211} (or similar) of the module
\url{special/advective_gauge.f90}.

The control parameters we use are the magnetic Reynolds number $\Rm$, the
magnetic Prandtl number $\Pm$, and the Schmidt number,
\begin{equation}
\Rm \equiv \frac{\urms}{\eta\kf}, \quad \Pm\equiv \frac{\nu}{\eta},
\quad \Sc\equiv \frac{\nu}{\kappa},
\end{equation}
where $\urms$ is the root mean square velocity.
We use $\kf=3k_1$ where $k_1$, the box wavenumber, is unity.
The numerical resolution is varied between $32^3$ and $256^3$ meshpoints
for values of $\Rey$ and $\Rm$ between 3 and 300.
In one case we used $\Rm\approx800$, which was only possible because in
that case we used $\Pm=10$, so that most of the energy gets dissipated
viscously, leaving relatively little magnetic energy at high wavenumbers
\citep{B11}.

\section{Importance of magnetic helicity density}

\subsection{Implications of \eq{h1} for dynamo theory}

Magnetic helicity is not only of interest by being a conserved quantity in ideal MHD, but
also by being the basis of a methodology to treat nonlinear helical MHD
dynamos, namely
dynamical $\alpha$ quenching \citep{KR82}.
This methodology relates the current helicity in small
scale fields with the magnetic helicity in small-scale fields, $\jj \cdot \bb \simeq k_f^2 \aaa
\cdot \bb$, and invokes the magnetic $\alpha$ effect \cite{PFL}.
The evolution
equation of the magnetic helicity density then becomes the evolution
equation of the magnetic part of the $\alpha$ effect and the nonlinear
evolution of the dynamo can be modeled.
This methodology has been
used successfully in systems where no net helicity flux is possible, and initial
work invoking the methodology has captured the behavior of at least one system with
finite helicity fluxes \cite{HubbardBrandenburg2010GAFD}.
A major prediction of the theory is that in the absence of preferential helicity
fluxes of small-scale fields, dynamo action is quenched
to sub-equipartition mean field strengths.
This phenomenon is sometimes referred to as ``catastrophic quenching''.

\subsection{Magnetic helicity as passive scalar}

In the advective and advecto-resistive gauges, the velocity appears in
the evolution equations of the magnetic helicity density,
\Eqs{h1}{dhar}, only as advection terms in the fluxes, \Eqs{h1flux}{FFFar} .
In the limit of ideal, incompressible, kinematic MHD,
\Eq{h1} is the evolution equation for a passive scalar.
Even in non-ideal MHD, if the fluctuations of $h^{\rm ar}$ due to the
velocity field $\UU$ were
purely advective in nature (i.e.\ passive), magnetic helicity transport
would only be resistive, large-scale advective, and/or turbulently diffusive.
This would forbid the preferential export of small-scale magnetic
helicity and might call for alternate solutions to the catastrophic
quenching problem than helicity fluxes \cite{ssHelLoss09}.

While in ideal MHD ($\eta=0$) the resistive terms in \eq{h1} vanish, resistive terms need not
vanish in the limit
of $\eta\to0$ (high $\Rm$).
For example, in a turbulent flow, Ohmic dissipation $\eta\JJ^2$
tends to a finite value as $\eta$ decreases.
The need for non-resistive solutions to the
build-up of magnetic helicity is therefore not a given.
We will examine this by performing kinematic simulations where the
Lorentz force is turned off, i.e.\ $c_{\rm L}=0$.

If the Lorentz force is significant, the fluctuations of
$h^{\rm ar}$ and $\UU$
might be correlated beyond simple turbulent diffusion concerns
(i.e.\ the fluctuations of $\har$ could
drive flow patterns).
In the limit of incompressible flows, if the helicity is uniform, then the only
source terms for helicity patterns of finite $k$ are the resistive terms.
The terms are small compared to dimensional
estimates for the velocity terms when $\Rm \gg 1$.
We will look for signals of magnetic helicity transport by examining
spectra of $\hr$ and $\har$ as (pseudo) scalars, together with spectra
of a true passive scalar.
As we will show, the advecto-resistive gauge is
adequately efficient at turbulently diffusing magnetic
helicity that no inertial range for the magnetic helicity density can be identified.  However,
the spectra of $\hr$ help elucidate previous 
results \cite{HubbardBrandenburg2010GAFD} which found diffusive fluxes,
but at values well below turbulent diffusivities.
Instead, our spectra show clear diffusive behavior in the inertial range,
but the mere existence of the inertial range implies non-diffusive behavior.

We emphasize that our spectra of $\hr$ and $\har$ have nothing to do with
the usual magnetic helicity spectrum that obeys a realizability condition
and whose integral gives the volume-averaged magnetic helicity.
Here we are looking instead at the power of the magnetic helicity density
as a (pseudo) scalar field.  Our $h_k$ measures the spatial variation of $h$.
In order to avoid confusion, we shall refer to these spectra as scalar spectra.

\section{Results}
\label{Results}

The results reported below for the magnetic helicity density $h$ refer
to the advecto-resistive gauge and have been obtained by the $\Lambda$
method, unless indicated otherwise.
The results from the direct method agree (\Sec{Agreement}), but this
method develops an instability when nonlinear effects become important
(\Sec{stability}).

\subsection{Agreement between $\Lambda$ and direct methods}
\label{Agreement}

\begin{figure}[t!]\begin{center}
\includegraphics[width=\columnwidth]{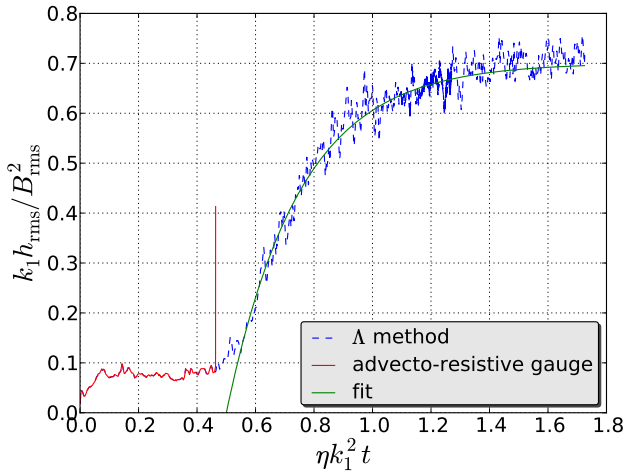}
\end{center}\caption[]{(Color online) Time dependence of the normalized helicity for the
advecto-resistive gauge with the direct method and the $\Lambda$ method.
Both curves agree perfectly just until the moment when the code develops
an instability in the direct calculation.
Time is normalized in terms of the magnetic diffusion time.
The fit is an exponential relaxation to a constant value proportional
to $1-\exp(-2\eta k_{\rm m}^2\Delta t)$, where $\Delta t=t-t_{\rm sat}$ is the
time after the small-scale magnetic field has saturated
\cite{Brandenburg2001ApJ} and $k_{\rm m}=1.4k_1$ has been chosen
for a good fit.
}\label{fig: h_departure}
\end{figure}

To test the agreement between the $\Lambda$ method and directly
solving the induction equation in the advecto-resistive gauge,
we plot the normalized rms magnetic helicity $h^{\rm ar}_{\rm rms}$
with respect to time (\Fig{fig: h_departure}).
Note that the non-dimensional ratio $k_1 h^{\rm ar}_{\rm rms}/B_{\rm rms}^{2}$
has a well-defined plateau during the kinematic stage.
Below we shall study the average value of this plateau as a function
of magnetic Reynolds and Prandtl numbers.
At the end of the kinematic phase, there is a slow saturation phase on
a resistive time scale during which the large-scale field of the dynamo
develops \cite{Brandenburg2001ApJ}.
The results of the two calculations agree just until the moment when
the direct calculation develops a numerical instability, whose nature
will be discussed in more detail below.
The perfect agreement until this moment can be taken as confirmation
that the $\Lambda$ method works and is correctly implemented in the code.

\begin{figure}[t!]\begin{center}
\includegraphics[width=\columnwidth]{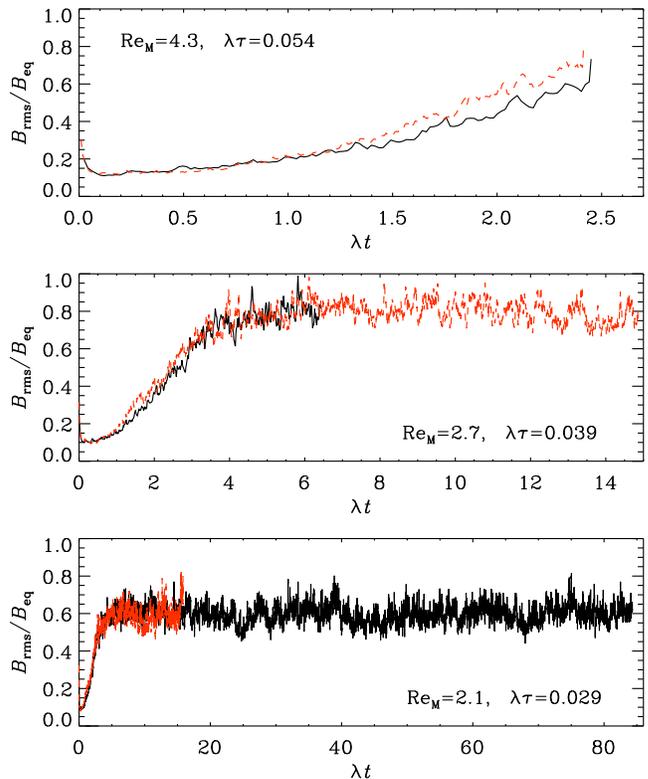}
\end{center}\caption[]{(Color online) Evolution of $B_{\rm rms}/B_{\rm eq}$ for small values
of $\Rm$ between $4.3$ (top) and 2.1 (bottom), using $32^3$ (solid lines)
and $64^3$ (dashed, red lines).
In each case, time on the abscissa is normalized by the growth rate
$\lambda$, whose value is given in each panel in units of the inverse
turnover time, $\tau^{-1}=\urms\kf$.
The ends of each line mark the point when the solution became unstable.
}\label{fig:pcomp32_64}
\end{figure}

\subsection{Nature of the instability}
\label{stability}

\begin{figure}[t!]\begin{center}
\includegraphics[width=\columnwidth]{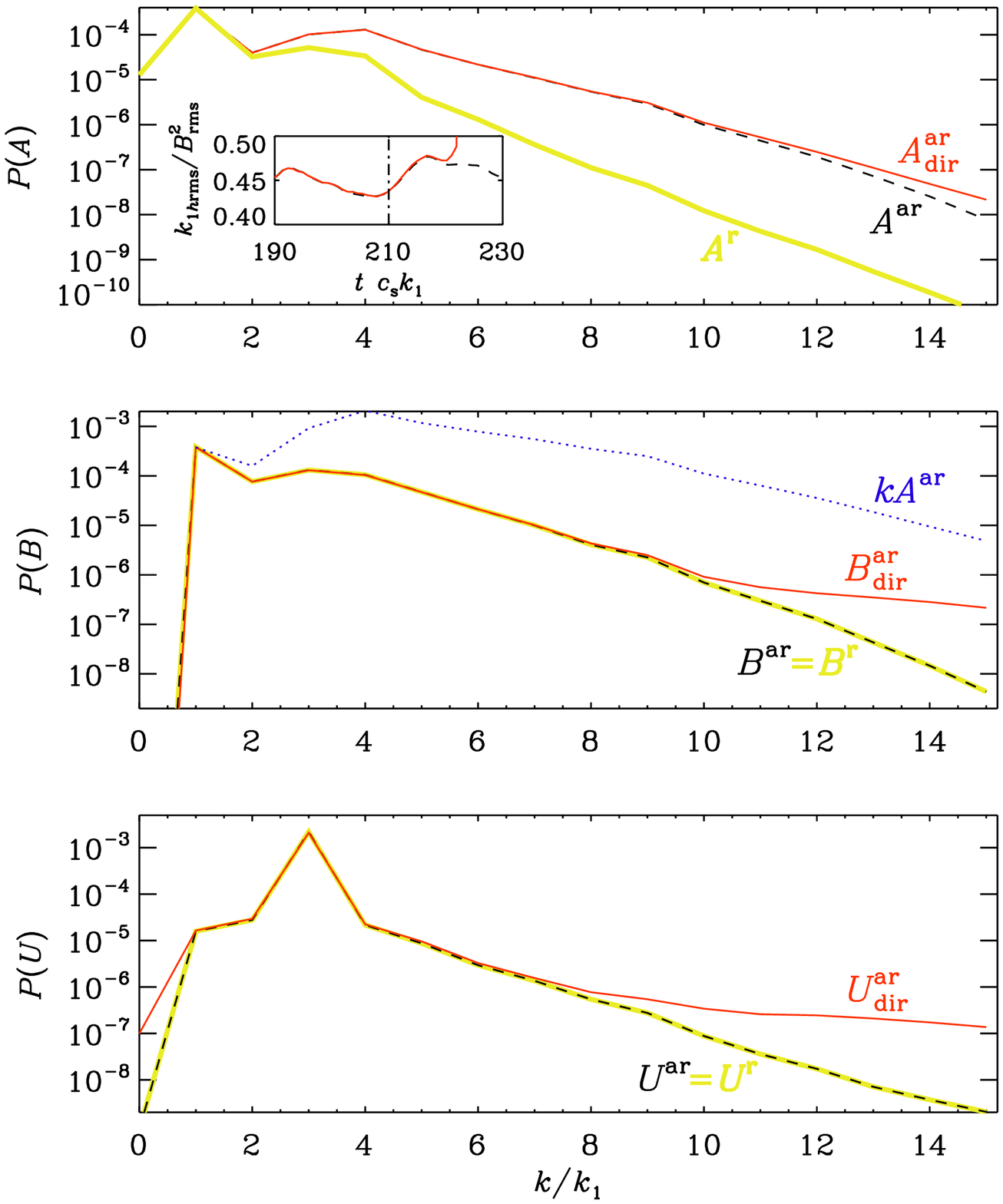}
\end{center}\caption[]{(Color online)
Power spectra of $\AAA$, $\BB$, and $\UU$ for two runs that are
identical except that the first run solves for $\AAAar$ directly
while the second solves for $\AAAr$ and $\Lambda$.
In the top panel we plot the spectrum of $\AAA$ obtained either via
$\AAAar=\AAAr+\nab \Lambda$ (dashed) or directly,
$\AAAar_{\rm dir}$ (solid/red), and compare with $\AAAr$ (thick gray/yellow),
showing that the vector potential
in the advecto-resistive gauge has much more power at high $k$.
The inset shows the time evolution of the normalized $h_{\rm rms}$ shortly
before the time of the numerical instability.
The dash-dotted line indicates the time for which the power spectra are taken.
In the second panel we present
magnetic energy spectra obtained in the direct gauge (solid/red), with the
$\Lambda$ method (dashed/black) as well as $k \AAAar$ (dotted/blue),
showing that there is significant power in the irrotational part of $\AAA$.
We see that in the direct calculation of $\AAAar$ the numerics is unable
to adequately handle
the high wavenumber power of $\AAAar$ with consequences for the velocity seen
in the last panel (solid/red line).
The spectra of $\BB$ and $\UU$ agree for resistive and advecto-resistive
gauges (thick gray/yellow line underneath the dashed black line) because the
evaluation of the curl of a gradient has been avoided (last two panels).
The three spectra are all taken for $t=210/\cs k_1$.
}\label{verification}
\end{figure}

In \Fig{fig:pcomp32_64} we show time series for a range of modest values
of $\Rm$ and two resolutions, $32^3$ and $64^3$.
Reducing the magnetic Reynolds number may
stabilize the system somewhat, but changing the resolution has no clear effect.
In \Fig{verification} we present data from equivalent runs that
solve either \eq{dAAAar} or alternatively \eqs{rdef1}{dLdt1}.  We can see
that the solutions match up until time $t=220/\cs\kf$, where the run that solves
\eq{dAAAar} becomes unstable.

The key point is that when we evolve \eqs{rdef1}{dLdt1}, $\Lambda$
never enters the equations for physical quantities.
However, when we evolve \eq{dAAAar}, the magnetic
field includes a term $\nab \times (\nab \Lambda)$, which,
when computed numerically,
is not zero.  
The first panel in \Fig{verification} shows the power spectra of the
vector potential.
Comparing the advecto-resistive gauge (dashed/red) with  resistive gauge (dotted/blue) 
we see that $\AAAar=\AAAr+\nab \Lambda$ has significantly more power at high $k$ than
$\AAAr$.  
Numerics cannot adequately handle the requirement that 
$\nab \times \nab \Lambda=0$ at high $k$ in the direct method, 
introducing errors in $\BB$, as can be seen in the second panel.  
This fictitious
increase in magnetic power at high $k$ (and the attendant increase in current) result in a
fictitious high $k$ increase in the velocity field (third panel) that produces the
numerical instability.
The results of
\Fig{fig:pcomp32_64} suggest that the power of $\Lambda$
(remembering that $\JJ$ includes that the third derivative of $\Lambda$) drops slowly enough at
high $k$ that numerical stability can only be achieved by enforcing an adequate resistivity $\eta$
to damp $\Lambda$ for only modest wavenumbers.
Indeed, any gauge with large power in $\AAA$ for high $k$ is expected to be numerically
unstable, and the method sketched in Appendix
\ref{sec: Derivation of Lambda} or \ref{sec: Derivation of dLdt1} may be
used to make the connection between analytical results in such a numerically unstable
gauge and numerical results produced in a stable gauge.

\subsection{Evolution of rms helicity density}
\begin{figure}[t!]\begin{center}
\includegraphics[width=\columnwidth]{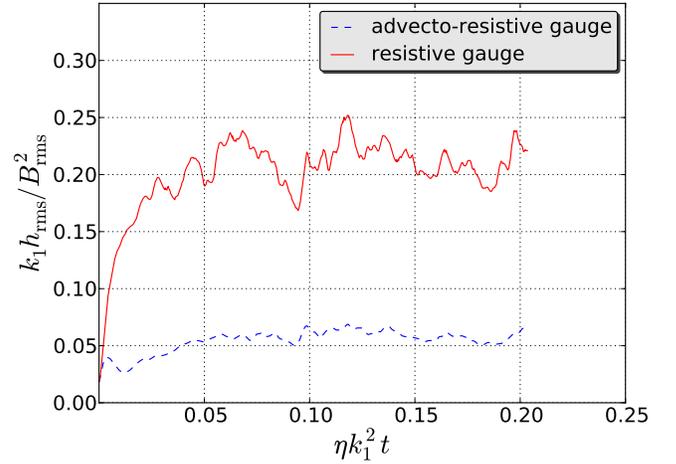}
\end{center}\caption[]{(Color online)
Time dependence of the rms values for the helicity in the advecto-resistive
(solid/red) and resistive (dashed/blue) gauges with the Lorenz force
switched off, i.e.\ $c_{\rm L}=0$ in both cases.}
\label{fig: hrms kinematic}
\end{figure}

In \Fig{fig: hrms kinematic} we present a time series of the normalized
rms magnetic helicity density in the kinematic regime
(Lorentz force turned off, i.e.\ $c_{\rm L}=0$).
In both the advecto-resistive and resistive gauges,
there is an initial adjustment of the non-dimensional ratio
$k_{1}h_{\rm rms}/B_{\rm rms}^{2}$ to a certain value, followed
by a plateau.
In the kinematic regime the magnetic helicity density is
passive and the advection term in the advecto-resistive gauge
merely serves to turbulently diffuse any local concentrations of $\har$.
Therefore there cannot be any spontaneous growth of $\har$, except for effects
from the resistive terms in the early adjustment phase.
Turbulent diffusion itself, on the other hand, cannot generate variance
of $\har$.

\begin{figure}[t!]\begin{center}
\includegraphics[width=\columnwidth]{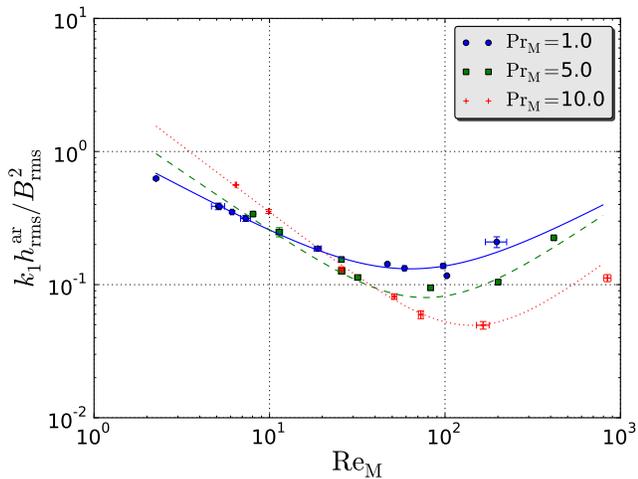}
\end{center}\caption[]{(Color online) 
$\Rm$ dependence of $k_{1}h^{\rm ar}_{\rm rms}/B_{\rm rms}^{2}$
for the kinematic phase. Values are averages over times where they reach a
stationary state. The curves represent fits according to \eqref{eq: fitfunc}.}
\label{har_brms2_Rm_kin}
\end{figure}

In \Figs{har_brms2_Rm_kin}{hr_brms2_Rm_kin}
we plot the height of the rms-magnetic helicity density plateau as a function
of $\Rm$
for several values of the magnetic Prandtl number and
constant forcing amplitude.
The differences between the evolution equations for $\hr$ and $\har$
are contained entirely
in the flux terms so the volume integral of $h$ is the same in the two gauges.
Any difference between the rms values of $h$ therefore is due to spatial fluctuations
generated by the flux terms.

\begin{figure}[t!]\begin{center}
\includegraphics[width=\columnwidth]{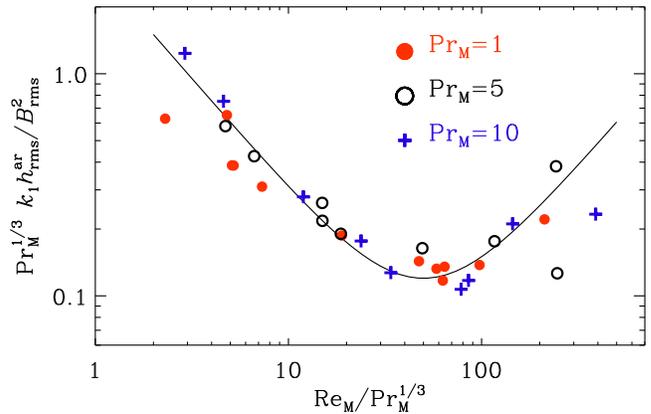}
\end{center}\caption[]{(Color online) 
Dependence of $k_{1}h^{\rm ar}_{\rm rms}/B_{\rm rms}^{2}$,
scaled by $\Pm^{1/3}$ on $\Rm/\Pm^{1/3}$ for the kinematic phase
and $\Pm=1$ (filled circles), 5 (open circles), and 10 (plus signs).
The solid line represents the fit of \Eq{eq: fitfunc2}.
}\label{pplateau_all_pm}
\end{figure}

We fit the data points in \Fig{har_brms2_Rm_kin}
with functions of the form
\EQ
\frac{k_1 \har}{B^2_{\rm rms}}=c\Rm^{-a}\left(1+b\Rm^{2a}\right).
\label{eq: fitfunc}
\EN
The fit results for the parameters are presented in \Tab{fits}.
Of interest is $c$, which increases with $\Pm$ and
scales approximately with
$\Pm^{1/2}$.
A more general, albeit less accurate fit is given by
\EQ
\frac{k_1 \har}{B^2_{\rm rms}}\approx3\,\Rm^{-1}
\left[1+\left({\Rm/\Pm^{1/3}\over50}\right)^2\right],
\label{eq: fitfunc2}
\EN
see \Fig{pplateau_all_pm}.

\begin{figure}[t!]\begin{center}
\includegraphics[width=\columnwidth]{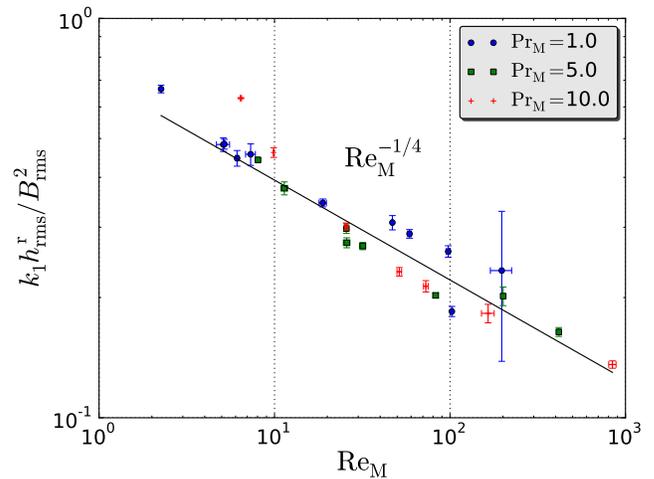}
\end{center}\caption[]{(Color online) 
$\Rm$ dependence of $k_{1}h^{\rm r}_{\rm rms}/B_{\rm rms}^{2}$
for the kinematic phase. Values are averages over times where they reach a
stationary state. A $-1/4$ power law can be seen.}
\label{hr_brms2_Rm_kin}
\end{figure}

\begin{figure}[t!]\begin{center}
\includegraphics[width=\columnwidth]{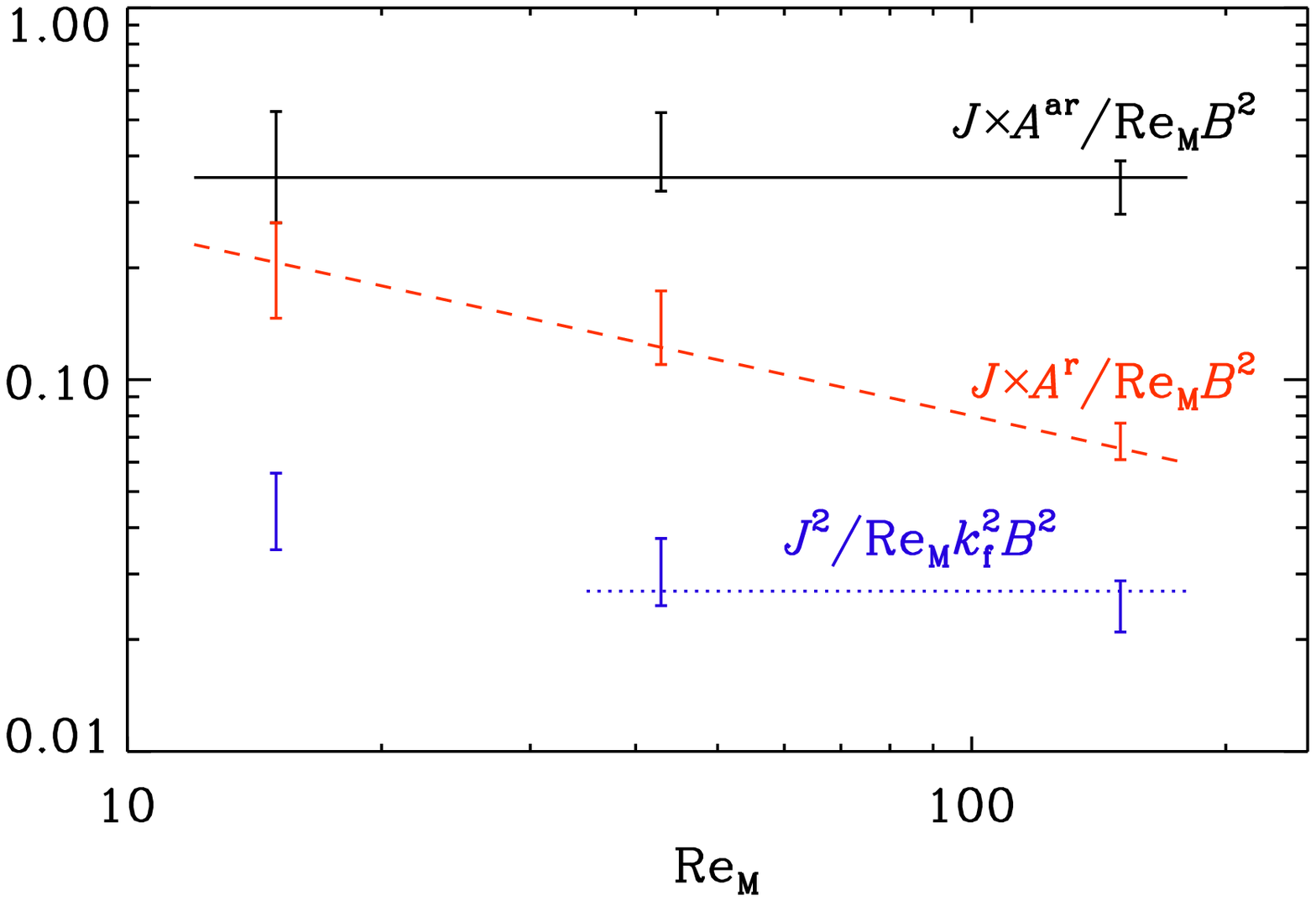}
\end{center}\caption[]{(Color online) 
$\Rm$ scaling of the rms value of $\JJ\times\AAA$, normalized by
$\Rm B_{\rm rms}^2$, for the advecto-resistive and resistive gauges.
The solid line represents constant scaling, i.e.\
$\eta\JJ\times\AAAar\approx\const$,
while the dashed line represents inverse square root scaling, i.e.\
$\eta\JJ\times\AAAr\propto\Rm^{-1/2}$,
for three runs with $\Pm=1$ in the saturated regime.
The dotted/blue line shows that $\eta J^2$, properly normalized,
is approximately constant.
}\label{ppn3}\end{figure}

It is clear that high wavenumber fluid eddies (which are damped
for small $\Rey$, i.e.\ large $\Pm$,
contribute significantly to $h^{\rm ar}_{\rm rms}$ for $\Rm>100$,
while from \Fig{hr_brms2_Rm_kin} we see that they do not contribute to
$h^{\rm r}_{\rm rms}$.
That these eddies could contribute in the advecto-resistive
gauge is to be expected as the advective nature of that gauge implies
the existence of an efficient turbulent cascade;
the fact that they do contribute there and that the $\eta\JJ\times\AAAar$
and $\eta(\nab\cdot\AAAar)\BB$ terms remain important implies 
that resistive terms both become important at small length scales and
have non-dissipative effects.
This is explained by the fact that $\AAAar$ develops a strong
high-$k$ tail; see also \Fig{verification}.
This is confirmed in \Fig{ppn3}, which shows that the resistive
magnetic helicity fluxes in the advecto-resistive gauge are proportional
to $\Rm$.
In this gauge the rms resistive helicity fluxes are therefore independent
of the actual value of the resistivity,
staying finite even in the high $\Rm$ limit.
This is quite different from the resistive magnetic helicity fluxes in the
resistive gauge, and the global magnetic helicity dissipation (which is
gauge-independent): both terms are only proportional to $\Rm^{1/2}$ and,
after multiplying with $\eta$ these terms tend to zero for $\Rm\to\infty$.

\begin{table}[b!]
\caption{Fit parameters for equation \eq{eq: fitfunc}
and \Fig{har_brms2_Rm_kin}.}
\label{fits}
\vspace{12pt}\centerline{\begin{tabular}{ccccl}
\hline \hline
$\; \Pm\; \; $ & $a $ & $ b $ & $ c $ & \; \; line type \\
\hline
$ 1 $ & $\; \; 0.7\; \;$ & $\; \; 3 \times 10^{-3}\; \;$ & $\; \; 1.2\; \; $ & \; \; solid/blue\\
$5$ & $0.9$ & $4 \times 10^{-4}$ & $2.0$ & \; \; dashed/green\\
$10$ & $1.0$ & $5 \times 10^{-5}$ & $3.5$ & \; \; dotted/red \\
\hline \hline
\end{tabular}}
\end{table}


\subsection{Comparison with passive scalar}

\begin{figure}[t!]\begin{center}
\includegraphics[width=\columnwidth]{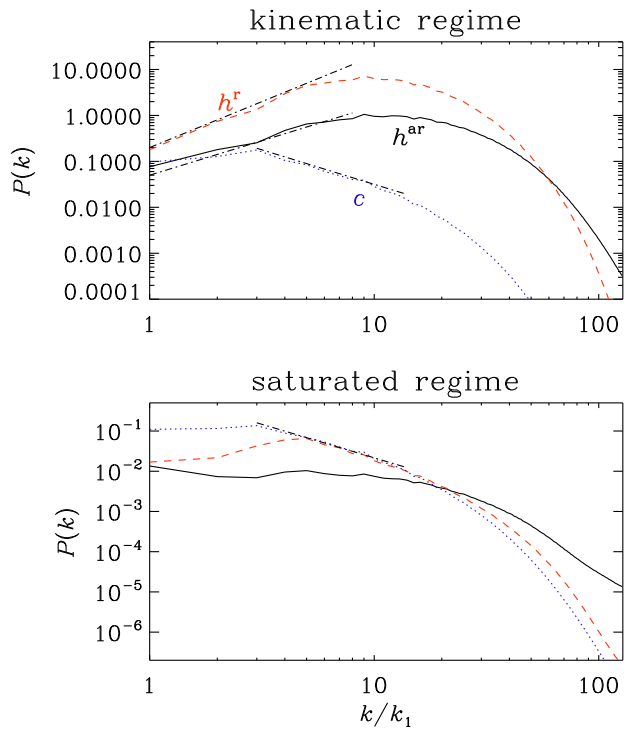}
\end{center}\caption[]{(Color online) 
Power spectra of $\hr$, $\har$, and the passive scalar $c$,
both in the kinematic regime (top)
and the nonlinear saturated regime (bottom) for $\Rey=80$ with $\Pm=\Sc=1$.
In the kinematic regime, the dash-dotted lines have slopes $+2$
for $\hr$, $+3/2$ for $\har$, and $-3/2$ for $c$ (top) and
$-5/3$ for $c$ in the saturated regime.
}\label{spectra}\end{figure}

In \Fig{spectra} we present scalar spectra of the magnetic helicity
density for both the resistive and advecto-resistive gauges and for the
passive scalar concentration $c$, in the kinematic (arbitrary units)
and saturated regimes.
The passive scalar spectrum shows a peak at the forcing scale,
$\kf/k_1=3$, followed by an approximate $k^{-5/3}$ subrange and
an exponential diffusive subrange.
As long as the magnetic energy density is still small compared with
the kinetic energy density, the field exhibits exponential growth
and a Kazantsev $k^{3/2}$ energy spectrum, which is well seen in
simulations even at magnetic Prandtl numbers of unity both with
and without kinetic helicity in the velocity field \cite{HBD04}.
This $k^{3/2}$ spectrum is also reflected in the scalar spectrum of $\har$.
The scalar spectrum of $\hr$ is somewhat steeper and closer to $k^2$,
indicating that $\hr$ is dominated by white noise in space at large scales.

The saturated regime exhibits some interesting properties.  The pronounced
peak of the power of the passive scalar at the driving scale is easily
understood as being due to the source of $c$.  However, the
magnetic helicity density in the resistive gauge shows a significant peak
there as well, while it does not
in the advecto-resistive gauge.  This implies that the
velocity term in \Eq{non-adv} generates significant spatial variations in the magnetic
helicity density -- even in the absence of external modulations.  As
in dynamical $\alpha$ quenching, $h$ influences the $\alpha$ effect, this
suggests a way to quantify the appropriateness of different gauge choices: systems
where spatial and temporal fluctuations in $\alpha$ can be adequately constrained
would allow one to determine whether spatial fluctuations in $h$, as seen
in \Fig{spectra}, are fictitious as suggested by the advecto-resistive
gauge or not.

The spectra of $\har$ in the saturated regime
does not present a clear inertial range, so we cannot draw strong conclusions
as to possible non-diffusive turbulent fluxes.
However,  $\hr$ follows the same
cascade as the passive scalar.
Previous studies in that gauge \cite{HubbardBrandenburg2010ApJ}
found that magnetic
helicity fluxes were best treated as diffusive, although the fits were imperfect.
The diffusive nature is clearly seen in the spectrum while the imperfections of the 
diffusive fit can be seen in the generation of a peak at the driving scale.
This evidence in support  of diffusive magnetic helicity
fluxes gives us the confidence to predict at what $\Rm$ diffusive magnetic helicity
fluxes will play a dominant role in dynamo saturation, i.e.\ when the diffusive
fluxes have a greater effect on magnetic helicity evolution than the resistive
terms.
This will be done in \Sec{Revisiting} where we re-analyze simulation data
from earlier work \cite{HubbardBrandenburg2010GAFD}.

\section{Revisiting earlier work}
\label{Revisiting}

Earlier work \cite{helGauge10,HubbardBrandenburg2010GAFD} on magnetic
helicity fluxes in inhomogeneous open systems confirmed that the magnetic
helicity density of the small-scale field is gauge-invariant -- even if
that of the large-scale field is not.
The divergence of the mean magnetic helicity flux of the small-scale field
is then also gauge-invariant, but its value is small compared with
resistive magnetic helicity dissipation.
We return to this work to estimate at what $\Rm$ diffusive magnetic helicity
fluxes will begin to play a dominant role in dynamo saturation.

\begin{figure}[t!]\begin{center}
\includegraphics[width=\columnwidth]{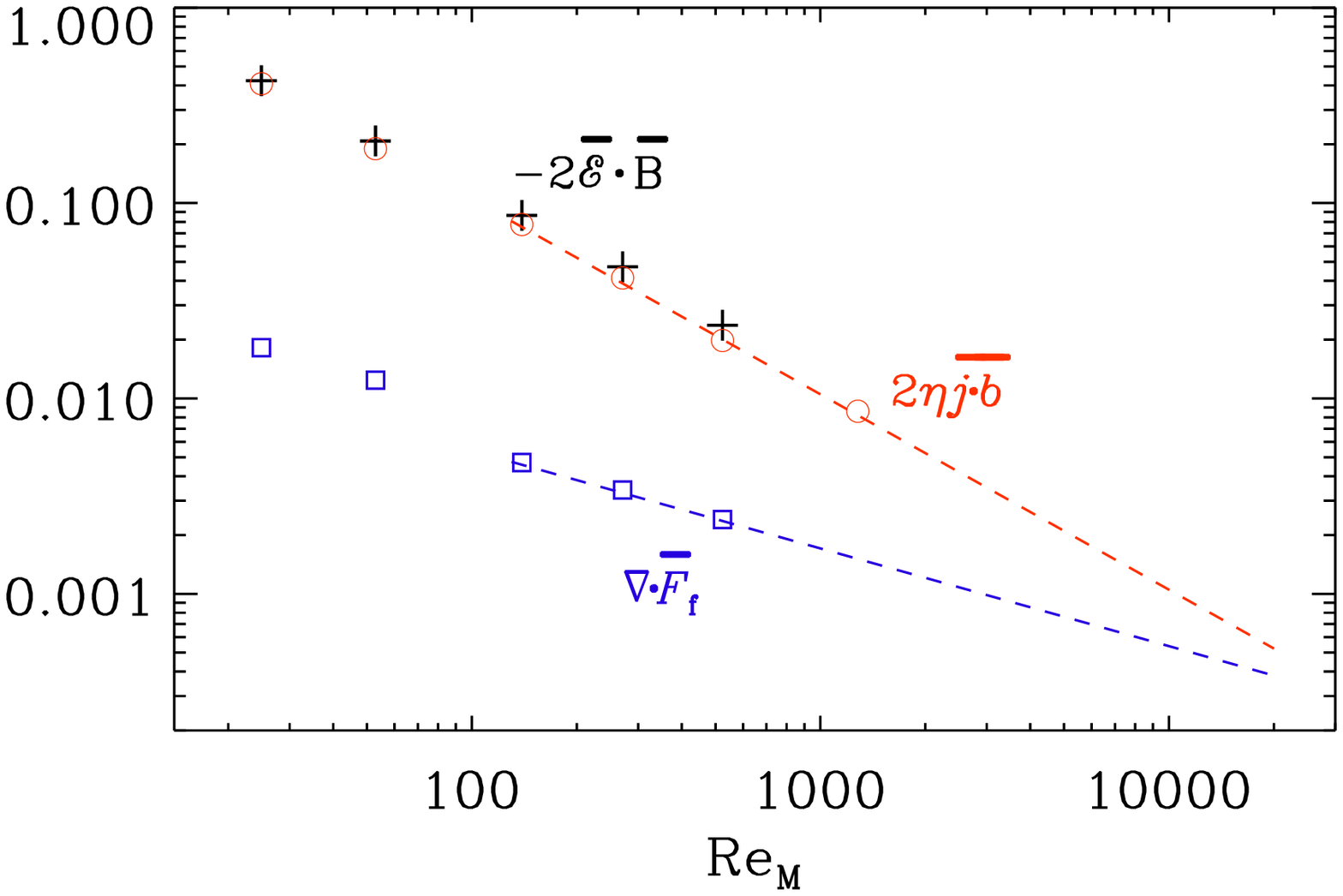}
\end{center}\caption[]{(Color online) 
Scaling of $\meanEMF\cdot\meanBB$, $\overline{\jj\cdot\bb}$, and
$\nab\cdot\FFFf$ versus $\Rm$ for the data of an earlier simulation
\cite{HubbardBrandenburg2010GAFD} of helically driven turbulence
embedded in a poorly conducting non-helically driven turbulent halo.
The symbols show actual data obtained from simulations, the 
dashed lines are the extrapolation to high $\Rm$. 
}\label{ppflux_with}\end{figure}

We emphasize that we are now discussing helicity properties of
what we call the small-scale field.
Such a field is defined by introducing an averaged magnetic field,
$\meanBB$, indicated by an overbar.
Following earlier work \cite{helGauge10,HubbardBrandenburg2010GAFD}
we restrict ourselves here to planar (or horizontal) averaging.
The small-scale field is then given by $\bb=\BB-\meanBB$, and the
mean magnetic and current helicity densities of the fluctuating fields
are then $\hf\equiv\overline{\aaa\cdot\bb}$ and $\overline{\jj\cdot\bb}$,
respectively, where $\nab\times\aaa=\bb$ and $\jj=\nab\times\bb$.
Turbulent diffusion and the $\alpha$ effect imply helicity transfer
between scales \cite{See96,Ji99}
through the mean electromotive force of the fluctuating
field, $\meanEMF=\overline{\uu\times\bb}$, so that the evolution
equation for $\hf$ takes the form
\EQ
{\partial\hf\over\partial t}=-2\meanEMF\cdot\meanBB
-2\eta\overline{\jj\cdot\bb}-\nab\cdot\FFFf.
\label{dhf}
\EN
Here, both $\hf$ and $\nab\cdot\FFFf$ are a gauge-dependent,
but if there is a steady state, and if $\hf$ is constant, then
$\partial\hf/\partial t=0$, and since both $\meanEMF\cdot\meanBB$ and
$\overline{\jj\cdot\bb}$ are gauge-invariant, $\nab\cdot\FFFf$ must
also be gauge-invariant.
Numerical values for $\meanEMF\cdot\meanBB$, $\overline{\jj\cdot\bb}$,
and $\nab\cdot\FFFf$ were given earlier \cite{HubbardBrandenburg2010GAFD}
for a particular simulation of a slab of helically driven turbulence
embedded in a poorly conducting non-helically driven turbulent halo.
In \Fig{ppflux_with} we show the scaling of all three terms versus $\Rm$.
Note that $-\meanEMF\cdot\meanBB$ is balanced mainly by
$\overline{\jj\cdot\bb}$.
However, if the current trend, $\overline{\jj\cdot\bb}\sim\Rm^{-1}$
and $\nab\cdot\FFFf\sim\Rm^{-1/2}$ were to continue, one might expect
a cross-over at $\Rm\approx3\times10^4$.
If so, the scaling of $\meanEMF\cdot\meanBB$ is expected
to become shallower, following that of  $\nab\cdot\FFFf$.
Given that the largest $\Rm$ accessible today is of order $10^3$,
we may conclude that an alleviation of quenching through
diffusive magnetic helicity fluxes will not be prominent in
simulations for the near future.  Nevertheless, astrophysical systems such
as the Sun are orders of magnitude beyond the estimated critical point of
$\Rm\sim 3\times 10^4$; and we expect their dynamo dynamics to behave
accordingly.

\section{Conclusions}
\label{Conclusions}

In view of the fact that the time averaged magnetic helicity of the
fluctuating fields is gauge-invariant in systems with sufficient
scale separation, the gauge-freedom can be exploited to gain insights
using gauges that are particularly revealing.
Here we have examined an interesting gauge, the advecto-resistive gauge.
As the advecto-resistive gauge
is inherently numerically unstable, we had to implement a possibly
universal technique to run numerical simulations in such
unstable gauges by running in a stable gauge while also solving a further
equation for the gauge transformation.

The advecto-resistive gauge has allowed us to examine both the
consequences of finite
resistivity for magnetic helicity density as well as the possibilities of
turbulent transport.
The magnetic helicity flux, and in particular the contribution from
$\eta\JJ\times\AAAar$ (properly normalized) reaches a constant value as $\eta\to 0$. This
behavior is similar to the behavior of energy dissipation in turbulence,
known as the law of finite energy dissipation \cite{Fri95}. 
This is interesting as the source term for
the volume integrated magnetic helicity $H$ does in fact tend to zero as $\eta$ does.
In this sense, the high $\Rm$ behavior of magnetic helicity is richer
than previously anticipated.
Indeed, the generation of spatial magnetic helicity fluctuations
\emph{ex nihilo} in non-advecto-resistive gauges is interesting, with
potentially testable implications.
We expect that the magnetic helicity fluxes resulting from terms of
the form $\eta\JJ\times\AAAar$ can be modeled as turbulent
Fickian diffusion-type fluxes down the gradient of mean magnetic helicity.
However, it is clear that fluxes from turbulent diffusion provide only
a poor escape from catastrophic $\alpha$ quenching, partly because
they cannot distinguish between large- and small-scale fields.
Furthermore, in simulations with such turbulent diffusion fluxes,
their contribution is still much smaller than the local resistive
magnetic helicity dissipation \cite{helGauge10,HubbardBrandenburg2010GAFD}.
However, the latter decreases faster ($\sim\Rm^{-1}$) with magnetic
Reynolds number than the former ($\sim\Rm^{-1/2}$), so one may estimate
that only for magnetic Reynolds numbers of
around $10^4$ one has a chance to see the effects of turbulent diffusion.
If true, however, such fluxes would definitely be important for the
magnetic Reynolds numbers relevant to stars and galaxies -- even
though such values cannot be reached with present day computer power.

\acknowledgements

National Supercomputer Centre in Link\"oping and the Center for
Parallel Computers at the Royal Institute of Technology in Sweden.
This work was supported in part by the Swedish Research Council,
grant 621-2007-4064, and the European Research Council under the
AstroDyn Research Project 227952.

\def\apj{Astrophys. J.}
\def\apjl{Astrophys. J. Lett.}
\def\mnras{Month. Not. Roy. Astron. Soc.}

\bibliographystyle{ieeetr}
\bibliography{references}

\appendix

\section{Derivation of \Eq{dLamWa}}
\label{sec: Derivation of Lambda}

We begin by expressing $\UU\times\BB$ in terms of $\AAA$,
\EQ
(\UU\times\BB)_i=U_jA_{j,i}-U_jA_{i,j}.
\EN
The last term can be subsumed into an advective derivative term for $\AAA$.
Using furthermore $U_jA_{j,i}=(U_jA_{j})_{,i}-U_{j,i}A_j$,
we can write \Eq{Weyl} as
\EQ
{{\DD\AW_i}\over\DD t}=-U_{j,i}\AW_j+(\UU\cdot\AAAW)_{,i}-\eta J_i.
\label{DDAW}
\EN
We now insert \Eq{LamWa} for $\AAAW=\AAAa-\nab\LamWa$, so
\EQA
{{\DD\Aa_i}\over\DD t}-{\DD\LamWa_{,i}\over\DD t}=\!\!\!\!&&
-U_{j,i}\Aa_j+U_{j,i}\LamWa_{,j}
\nonumber \\
&&+(\UU\cdot\AAAW)_{,i}-\eta J_i.
\label{AaLamWa}
\ENA
and note that
\EQ
-{\DD\LamWa_{,i}\over\DD t}=-\nabla_i\left({\DD\LamWa\over\DD t}\right)
+U_{j,i}\LamWa_{,j}.
\EN
The last term cancels and we are left with
\EQ
{{\DD\Aa_i}\over\DD t}+U_{j,i}\Aa_j+\eta J_i=
\nabla_i\left({\DD\LamWa\over\DD t}+\UU\cdot\AAAW\right),
\EN
so we recover the evolution equation for the advective gauge
provided \Eq{dLamWa} is obeyed.

\section{Derivation of \Eq{dLdt1}}
\label{sec: Derivation of dLdt1}

We present here the derivation of the transformation from the resistive
gauge to the advecto-resistive gauge, proceeding analogously to the
derivation presented in \App{sec: Derivation of Lambda}.
However, instead of \Eq{DDAW} we now have
\EQ
{{\DD\Ar_i}\over\DD t}=-U_{j,i}\Ar_j
+(\UU\cdot\AAAr)_{,i}+\eta\nabla^2\Ar_i.
\label{DDAr}
\EN
Inserting \Eq{Lamrar} for $\AAAr=\AAAar-\nab\Lamrar$, we obtain
an Equation similar to \eq{AaLamWa}, 
\EQA
{{\DD\Aar_i}\over\DD t}-{\DD\Lamrar_{,i}\over\DD t}=\!\!\!\!&&
-U_{j,i}\Aar_j+U_{j,i}\Lamrar_{,j}+(\UU\cdot\AAAr)_{,i}
\nonumber \\
&&+\eta\nabla^2\Aar_i-\eta\nabla^2\Lamrar_{,i}.
\label{ArLamrar}
\ENA
which leads to
\EQA
{{\DD\Aar_i}\over\DD t}&&+U_{j,i}\Aar_j-\eta\nabla^2\Aar_i=
\nonumber \\
&&\nabla_i\left({\DD\Lamrar\over\DD t}+\UU\cdot\AAAr-\eta\nabla^2\Lamrar\right),
\ENA
so we recover the evolution equation for the advecto-resistive gauge
provided \Eq{dLdt1} is obeyed.

\end{document}